\documentclass{ieeetj}
\usepackage{cite}
\usepackage{amsmath,amssymb,amsfonts}
\usepackage{algorithmic}
\usepackage{graphicx,color}
\usepackage{textcomp}
\usepackage{xcolor}
\usepackage{hyperref}
\hypersetup{hidelinks=true}
\usepackage{algorithm,algorithmic}

\usepackage{here}
\usepackage{color}
\usepackage{url}
\usepackage{amsxtra}
\usepackage{threeparttable}
\usepackage{comment}
\usepackage{upgreek}
\usepackage{multirow}
\usepackage{url}
\usepackage{booktabs}
\usepackage{setspace}
\usepackage{adjustbox}

\def\BibTeX{{\rm B\kern-.05em{\sc i\kern-.025em b}\kern-.08em
    T\kern-.1667em\lower.7ex\hbox{E}\kern-.125emX}}
\AtBeginDocument{\definecolor{tmlcncolor}{cmyk}{0.93,0.59,0.15,0.02}\definecolor{NavyBlue}{RGB}{0,86,125}}

\newcommand{\argmin}{\mathop{\rm argmin}\limits}
\def\x{{\mathbf x}}
\def\y{{\mathbf y}}
\def\X{{\mathbf X}}

\def\w{{\mathbf w}}
\def\b{{\mathbf b}}

\def\R{{\mathbb R}}

\def\z{{\mathbf z}}
\def\r{{\mathbf r}}
\def\b{{\mathbf b}}

\def\v{{\mathbf v}}
\def\A{{\mathbf A}}
\def\X{{\mathbf X}}
\def\W{{\mathbf W}}
\def\I{{\mathbf I}}
\def\Phi{\boldsymbol{\Upphi}}
\DeclareMathOperator{\prox}{prox}

\DeclareMathOperator{\TE}{TE}

\DeclareMathOperator{\DR}{DR}
\DeclareMathOperator{\MDTE}{MDTE}

\DeclareMathOperator{\diag}{diag}

\def\OJlogo{\vspace{-4pt}$<$Society logo(s) and publication title will appear here.$>$}
\def\seclogo{\vspace{10pt}$<$Society logo(s) and publication title will appear here.$>$}

\def\authorrefmark#1{\ensuremath{^{\textbf{#1}}}}

\begin{document}
\bstctlcite{IEEEexample:BSTcontrol}
\receiveddate{XX Month, XXXX}
\reviseddate{XX Month, XXXX}
\accepteddate{XX Month, XXXX}
\publisheddate{XX Month, XXXX}
\currentdate{XX Month, XXXX}
\doiinfo{XXXX.2022.1234567}

\markboth{}{Author {et al.}}

\title{Sparse Index Tracking: Simultaneous Asset Selection and Capital Allocation via $\ell_0$-Constrained Portfolio}

\author{Eisuke Yamagata\authorrefmark{1}, Student Member, IEEE, Shunsuke Ono. Author\authorrefmark{1}, Member, IEEE}
\affil{Tokyo Institute of Technology, Tokyo, Japan}
\corresp{(email: yamagata.e.ab@m.titech.ac.jp), (email: ono@c.titech.ac.jp).}
\authornote{This work was supported Grant-in-Aid for JSP Fellows 23KJ0915, in part by JST PRESTO under Grant JPMJPR21C4 and JST AdCORP under Grant JPMJKB2307, and in part by JSPS KAKENHI under Grant 22H03610, 22H00512, and 23H01415.}

\begin{abstract}
  Sparse index tracking is a prominent passive portfolio management strategy that constructs a sparse portfolio to track a financial index. A sparse portfolio is preferable to a full portfolio in terms of reducing transaction costs and avoiding illiquid assets. 
  To achieve portfolio sparsity, conventional studies have utilized $\ell_p$-norm regularizations as a continuous surrogate of the $\ell_0$-norm regularization. 
  Although these formulations can construct sparse portfolios, their practical application is challenging due to the intricate and time-consuming process of tuning parameters to define the precise upper limit of assets in the portfolio.
   In this paper, we propose a new problem formulation of sparse index tracking using an $\ell_0$-norm constraint that enables easy control of the upper bound on the number of assets in the portfolio. 
   Moreover, our approach offers a choice between constraints on portfolio and turnover sparsity, further reducing transaction costs by limiting asset updates at each rebalancing interval.
    Furthermore, we develop an efficient algorithm for solving this problem based on a primal-dual splitting method. Finally, we illustrate the effectiveness of the proposed method through experiments on the S\&P500 and Russell3000 index datasets.
\end{abstract}

\begin{IEEEkeywords}
$\ell_0$-norm constraint, Primal-dual splitting, Sparse index tracking.
\end{IEEEkeywords}


\maketitle

\section{Introduction}
\IEEEPARstart{I}{nvestors} can follow two basic approaches, namely, active and passive investment \cite{passive,active}.
Active investment strategies seek to outperform the market through active and strategic trading, aiming at short-term profits, whereas passive strategies strive to replicate market index performances, predicated on the belief that surpassing the market in the long term is unattainable \cite{barber2000trading}. As the stock market has historically grown, one can expect reasonable returns just by tracking its performance. This motivates investors to adopt passive over active strategies, considering how active strategies are riskier compared to passive strategies.

Ideally, trading an index directly would guarantee an investor returns consistent with that of the index.
However, because a financial index comprises multiple assets (e.g., 500 in the case of the S\&P500), direct investment in an index is not feasible for an investor.
An investor is required to invest in individual assets, which raises two key questions: 1) In which assets should the investor invest? And 2) How much should the investor invest in each of the assets?

To answer these questions, index tracking has been researched as one of the prominent passive investment strategies that tracks the performance of a market index by constructing a portfolio, which determines the target assets and the investment percentage of the assets. 
In theory, a portfolio that allocates appropriate fractions of the capital based on the benchmark weights that the index sponsor actually uses should perfectly track the target index.
However, this is not practically feasible since the benchmark weights are not publicly disclosed and are considerably expensive to acquire.
Another straightforward approach is to evenly distribute capital across all assets composing the index, effectively creating a fully and uniformly weighted portfolio.
However, this strategy has notable disadvantages: firstly, it necessitates substantial trading with each rebalance, inflating transaction costs; secondly, it incorporates small and illiquid assets, thereby increasing risk.
Given these disadvantages, \textit{sparse portfolios} with a limited maximum number of assets in the portfolio are much more desirable than full or dense portfolios.

\vspace{-2.5mm}
\subsection{Related Work}
The objective of sparse index tracking \cite{GAIVORONSKI2005115,8064704,Benidis_2018,9053677,9775642} is to construct a sparse portfolio $\w \in \R^{N}$, matching the performance of the benchmark market index, with $K(<<N)$ nonzero weights out of $N$ total assets.

The conventional approach to this was to divide the problem into two phases, namely, asset selection and capital allocation. Numerous methods for asset selection have been proposed. A typical approach was to select $K$ largest assets in terms of market capitalization \cite{OH2005371}. Other methods include selecting assets that exhibit similar performance to the target index \cite{DOSE2005145,bianchi} or selection based on the cointegration between log-prices of the $K$ assets and the index value \cite{carol}.
However, the impact of the two-step approach on tracking performance remained uncertain, prompting the development of one-step methods \cite{8064704,Benidis_2018} as alternatives.
The LAIT (Linear Approximation for Index Tracking) algorithm proposed in \cite{8064704} confirmed the positive effects of simultaneous asset selection and capital allocation by solving the following regression problem:
\begin{align}
  \label{eq:index_tracking}
  \begin{gathered}
   \min_{\w} \frac{1}{T}\|\r^\b - \X\w\|^2_2 + \lambda \mathrm{R}(\w),\;
   \mbox{s.t.} 
   \begin{cases}
    \w \geq \mathbf0,\\
    \mathbf1^\top \w = 1.
   \end{cases}
  \end{gathered}
\end{align}
where $\mathrm{R}(\w)$ is set to an $\ell_p$-norm regularization ($0<p<1$)\footnote{\setstretch{0.75}Although $\|\cdot\|_p$ is strictly speaking not a norm when $0\leq p<1$, it is conventionally referred to as the $\ell_p$-norm.} to enforce portfolio sparsity\footnote{\setstretch{0.75}To avoid nonconvex optimization, the $\ell_1$-norm is commonly used as a surrogate function of the $\ell_0$-norm. However, due to the short-selling and sum-to-one constraints, the $\ell_1$-norm of the portfolio is always a constant ($\|\w\|_1 = 1$). Therefore, the $\ell_1$-norm regularization cannot be used in this formulation.}.
The benchmark index returns across $T$ days are denoted by $\r^\b = [r^\b_1,...,r^\b_T]^\top \in \R^{T}$ and the asset-wise returns across $T$ days are denoted by $\X = [\r_1,...,\r_T]^\top \in \R^{T\times N}$. The nonnegative constraint prohibits short-selling (negative weights), and the sum-to-one constraint is enforced to allocate weights based on a constant capital. 

However, in the formulation of (\ref{eq:index_tracking}), there is no explicit relationship between the hyperparameter $\lambda$ and the number of assets in the portfolio, making it difficult to search for a $\lambda$ that would maintain a certain number of assets. This means that the investor cannot directly set the desired number of assets, which hinders actual asset management.

To solve this, an algorithm named NNOMP-PGD \cite{9775642} that combines nonnegative orthogonal matching pursuit \cite{7012095} and projected gradient descent \cite{8847410} has been proposed. The method separates the sparse index tracking problem into two stages:
\begin{align}
  \label{eq:selection}
  \begin{gathered}
  \min_{\w} \frac{1}{T}\|\r^\b - \X\w\|^2_2,\;
  \mbox{s.t.} 
  \begin{cases}
    \|\w\|_0 \leq K,\\
    \w \geq \mathbf0,
  \end{cases}
  \end{gathered}
\end{align}
and
\begin{align}
  \label{eq:allocation}
  \begin{gathered}
  \min_{\w} \frac{1}{T}\|\r^\b - \X\w\|^2_2,\;
  \mbox{s.t.}
  \begin{cases}
    \w \geq \mathbf0,\\
    \mathbf1^\top \w = 1.
  \end{cases}
  \end{gathered}
\end{align}
The first stage solves the asset selection problem with an $\ell_0$-norm \cite{7178689,Guoyin2015,L0,9970585,10138053,10251973} constraint (i.e., the number of nonzero entries in $\w$ is less than or equal to a user-set parameter) by NNOMP, and the second stage solves the capital allocation problem by PGD. By solving a formulation enforced with an $\ell_0$-norm constraint, investors can easily control the number of assets that compose the portfolio by the parameter $K$. 

Although NNOMP-PGD does succeed in directly controlling the sparsity of the portfolio through the $\ell_0$-norm constraint, we believe that the tracking performance can be improved by conducting the asset selection and capital allocation simultaneously.
This is due to the nonnegative orthogonal matching pursuit algorithm, wherein the value of the nonzero element added to $\w$ per iteration directly influences the selection of the nonzero element in $\w$ for the following iterations.
The nonzero element values do not necessarily satisfy the sum-to-one constraint or even the upper limit. 
\textit{This leads us to the question: Can we devise an $\ell_0$-norm constraint-based sparse index methodology that method simultaneously handles asset selection and capital allocation, thereby enhancing tracking performance?}

\vspace{-2mm}
\subsection{Contributions and Paper Organization}

This paper introduces an algorithm leveraging the primal-dual splitting method (PDS) \cite{pds}, designed to approximately solve sparse index tracking problems that address an $\ell_0$-norm constraint. This approach uniquely integrates asset selection and capital allocation without dividing them into separate stages.
Moreover, we extend the traditional sparse index tracking framework to (i) offer a choice between constraints on portfolio and turnover sparsity\footnote{\setstretch{0.75}A turnover constraint is considered in \cite{8064704,Benidis_2018}, wherein sparsity is enforced by an $\ell_1$-norm regularization. We, on the other hand, impose an $\ell_0$-norm constraint for a direct control of the sparsity.}, (ii) incorporate different tracking error metrics, and (iii) implement a box constraint (replacing the nonnegative constraint) to establish an upper limit on investments, mitigating risks associated with extreme capital distributions.
Through experiments on the S\&P500 and Russell3000 index datasets, we demonstrate the advantages of simultaneous asset selection and capital allocation and the positive effects of the generalized formulation.

The main contributions of this paper are as follows.
\begin{itemize}
  \item We develop a PDS-based algorithm that can directly handle an $\ell_0$-norm constraint while performing asset selection and capital allocation simultaneously.
  \item We generalize the conventional sparse index tracking formulation so that either portfolio or turnover sparsity can be enforced.
\end{itemize}

The paper is organized as follows. In Section \ref{sec:preliminary}, we introduce some basic concepts related to sparse index tracking and the mathematical instruments we will utilize in the proposed method. In Section \ref{sec:proposal}, we present the proposed method, along with its benefits and optimization algorithm. Finally, in Section \ref{sec:experiments}, we evaluate our method on real-world datasets.

\vspace{-3mm}
\section{Preliminary}
\label{sec:preliminary}

\subsection{Tracking Error Measure}
\label{subsec:TE}
 Several measures of tracking error (TE) have been proposed, the most common measure being the empirical tracking error (ETE) \cite{Maringer2007,BEASLEY2003621,Jansen2002,8064704}:
 \begin{align}
  \mathrm{ETE(\w)} := \frac{1}{T}\|\r^\b - \X\w\|^2_2.
 \end{align}
The empirical tracking error is convex and differentiable as
\begin{align}
  \nabla\mathrm{ETE(\w)} = -\frac{2}{T}\X^\top(\r^\b - \X\w).
\end{align}
The gradient $\nabla\mathrm{ETE(\w)}$ is $\beta$-Lipschitz continuous, where $\beta = 2/T\lambda_1(\X^\top\X)$ ($\lambda_1(\cdot)$ denotes the maximum eigenvalue of $\cdot$).

 Considering the original purpose of index tracking, which is to make a profit by investing, ETE that penalizes even when the portfolio beats the benchmark index is not desirable. Therefore, the measure downside risk (DR) \cite{GAIVORONSKI2005115,8064704} has been proposed to penalize only when the returns are behind that of the benchmark index:
 \begin{align}
  \mathrm{DR(\w)} := \frac{1}{T}\|(\r^\b - \X\w)^+\|^2_2,
 \end{align}
 where $[(\r^\b - \X\w)^+]_i = \max\{[\r^\b - \X\w]_i,0\}$. 
The downside risk is also convex and differentiable.
Consider $g_1(\w') = \|(\w')^+\|_2^2/T$ and $g_2(\w) = \r^\b - \X\w$ so that $\DR = g_1(g_2(\w))$.
The derivatives are
\begin{align}
  \begin{split}
  [\nabla g_1(\w')]_i &= 
  \begin{cases}
    2\w'_i/T,  &\text{if } \w'_i \geq 0,\\
   0,  &\text{otherwise},
\end{cases}\\
\therefore \nabla g_1(\w') &= 2(\w')^+/T,
\end{split}
\end{align}
and
\begin{align}
  \nabla g_2(\w) = -\X^\top,
\end{align}
respectively.
Therefore:
\begin{align}
  \begin{split}
  \nabla \DR(\w) &= \nabla (g_1 \circ g_2)(\w)\\
  &= \nabla g_2(\w)\nabla g_1(g_2(\w))\\
  &=-\frac{2}{T}\X^\top (\r^b - \X\w)^+.
  \end{split}
\end{align}
Furthermore, when comparing ETE and DR, the gradient of DR clearly does not exceed that of ETE. Therefore, DR can be considered $\beta$-Lipschitz continuous for at least the same value of $\beta$ as ETE.

\vspace{-3mm}
\subsection{Transaction Costs}

The typical transaction costs model applied in the U.S. markets is $\$0.005 \times v$ with a minimum cost of $\$1$, where $v$ is the trading volume. When the total capital is sufficiently large (making the volume-wise cost predominant), the total transaction cost is not influenced by the number of assets traded. Conversely, with smaller capital amounts, where the minimum cost prevails, the number of traded assets directly impacts transaction costs. Thus, a sparse portfolio is desired. 

In the same context, a turnover constraint \cite{8064704,Benidis_2018} has been introduced to cap the number of assets traded during each rebalancing period, defined as $\|\w_0 - \w\|_0\leq K$,
 with $\w_0$ representing the portfolio prior to rebalancing. By imposing this constraint, only a limited number of assets are updated, limiting the number of assets traded, therefore reducing transaction costs.

 \vspace{-3mm}
 \subsection{Proximal Tools}

The proximity operator \cite{moreau1962dual} of index $\gamma > 0$ of a proper lower semicontinuous convex function $f \in \Gamma_0(\R^n)$\footnote{\setstretch{0.75}The set of all proper lower semicontinuous convex functions on $\R^n$ is
denoted by $\Gamma_0(\R^n)$.} is defined as
\begin{align}
  \begin{gathered}
  \prox_{\gamma f}(\x):=\argmin_{\y} f(\y) + \frac{1}{2\gamma}\|\y-\x\|^2_2.
  \end{gathered}
\end{align}

The indicator function of a nonempty closed convex set $C$, denoted by $\iota_C$, is defined as
\begin{align}
  \iota_C(\mathbf{x}):=
  \begin{cases}
  0, \:\mathrm{if} \ \mathbf{x} \in C,\\
  \infty,\:\mathrm{otherwise}.
  \end{cases}
 \end{align}
 Since the function returns $\infty$ when the input vector is outside of $C$, it acts as the hard constraint represented by $C$ in minimization. The proximity operator of $\iota_C$ is the metric projection onto $C$, given by 
 \begin{align}
  \begin{gathered}
 \prox_{\iota_C}(\mathbf{x}) = P_C(\mathbf{x}) := \argmin_{\y \in C} \|\y -\x\|_2.
\end{gathered}
\end{align}

\subsection{Primal-Dual Splitting Method}
\label{sec:tools}
A primal-dual splitting method (PDS) \cite{pds,vu2013splitting}\footnote{\setstretch{0.75}This algorithm is a generalization of the primal-dual hybrid gradient method \cite{chambolle2011first}.} can solve optimization problems in the form of
\begin{align}
  \label{eq:pds}
  \begin{gathered}
    \min_{\w} f_1(\w) + f_2(\w) + f_3(\mathbf{A\w})\;  \mbox{s.t.} \;\mathbf{A\w} = \mathbf{v}
  \end{gathered}
\end{align}
where $f_1$ is a differentiable convex function with the $\beta$-Lipschitzian gradient $\nabla f_1$ for some $\beta > 0$, the proximity operators of $f_2 \in \Gamma_0(\R^n)$ and $f_3 \in \Gamma_0(\R^m)$ are efficiently computable (proximable), and $\mathbf{A} \in \R^{m \times n}$ is a matrix. The auxiliary variable $\mathbf{v}$ is used for the update in the following algorithm.
The problem (\ref{eq:pds}) is solved by the algorithm:
\begin{eqnarray}
  \label{eq:al_pds}
  \begin{gathered}
    \begin{split}
    \w^{(i+1)} &= \prox_{\gamma_1 f_2}{[\w^{(i)} - \gamma_1 (\nabla f_1 (\w^{(i)}) + \mathbf{A}^\top \mathbf{v}^{(i)})]},\\
    \mathbf{v}^{(i+1)} &= \prox_{\gamma_2 f_{3}^{*}}{[\mathbf{v}^{(i)} + \gamma_2 \mathbf{A}(2 \w^{(i+1)} - \w^{(i)})]},
    \end{split}
  \end{gathered}
\end{eqnarray}
where $f_{3}^{*}$ is the Fenchel-Rockafellar conjugate function of $f_3$ and the stepsizes $\gamma_1,\gamma_2>0$ satisfy $\frac{1}{\gamma_1} - \gamma_2 \lambda_1 (\mathbf{A}^\top\mathbf{A}) \geq \frac{\beta}{2}$.
The proximity operator of $f^{*}$ can be stated as the following \cite[Remark 14.4]{bauschke2017}:
\begin{align}
  \prox_{\gamma f^{*}}(\x) = \x - \gamma \prox_{\gamma^{-1}f}(\gamma^{-1}\x).
\end{align}
When the optimization problem is convex, the sequence $(\w^{(k)})_{k \in \mathbb{N}}$ theoretically converges to a solution of (\ref{eq:pds}) under some mild conditions on $f_2$, $f_3$, and $\mathbf{A}$ \cite[Theorem 3.1-3.3]{pds}. 

PDS has played a central role in various signal estimation methods, e.g., \cite{condat2014generic,ono2014hierarchical,ono2017primal,boulanger2018nonsmooth,kyochi2021epigraphical,9834326,9721064,yamagata2023robust}. A comprehensive review on PDS can be found in \cite{komodakis2015playing,20M1379344}.

\vspace{-5mm}
\section{Proposed Method}
\label{sec:proposal}
\subsection{Problem Formulation}

We formulate a new $\ell_0$-norm based index tracking problem that allows the selection of portfolio and turnover sparsity constraints.

The formulation is as follows:
\begin{align}
  \label{eq:ours}
  \begin{gathered}
   \min_{\w} \TE{(\w)},\;
   \mbox{s.t.}
   \begin{cases}
    \w \in S_s,\\
    \w \in S_{l,u},\\
    \mathbf1^\top \w = 1.
   \end{cases}
  \end{gathered}
\end{align}
Here, $\w$ represents the portfolio weights, $S_s$ embodies the sparsity constraints, and $S_{l,u}$ represents the box constraints.
The tracking error, denoted as $\TE$ is either ETE or DR introduced in Section \ref{sec:preliminary}-\ref{subsec:TE}.
The sparsity constraint is denoted by $S_s = S_0$ or $S_s = S_{w_0}$ where 
\begin{eqnarray}
  \begin{gathered}
    \begin{split}
S_0&:=\{\w\in\R^N \mid  \|\w\|_0\leq K_1\} \\
S_{\w_0}&:=\{\w\in\R^N \mid  \|\w - \w_0\|_0\leq K_2\}. 
\end{split}
\end{gathered}
\end{eqnarray}
When $S_s = S_0$, the formulation imposes sparsity on the portfolio in order to output a portfolio of $K_1$ sparseness.
On the other hand, when $S_s = S_{\w_0}$, the sparsity applies to the turnover, thus producing a portfolio that only requires $K_2$ trades from the previous portfolio ($\w_0$).
Note that although $S_0$ is a special case of $S_{\w_0}$ when $\w_0=\mathbf0$, we distinguish the two for clarity and to differentiate $K_1$ and $K_2$.

The set $S_{l,u}$ is a box constraint with a lower bound of $l$ and an upper bound of $u$. The lower bound is set to $l=0$ to prohibit short selling, and an upper bound is set to avoid extreme capital allocations, which is often risky in investment. 
\vspace{-5mm}
\subsection{Algorithm}

\begin{algorithm}[t]
  \caption{PDS-based algorithm for solving (\ref{eq:ours})}
  \label{al:ours1}
  \begin{algorithmic}[1]
  \renewcommand{\algorithmicrequire}{\textbf{Input:}}
  \renewcommand{\algorithmicensure}{\textbf{Output:}}
  \REQUIRE $\r^\b$,  $\X$
  \ENSURE Output signal $\w$ \\
  \textbf{Initialize $\w = \mathbf0$}
  \WHILE{{\it A stopping criterion is not satisfied}}
   \STATE{$\w^{(k+1)} \!\leftarrow\! P_{S_s}\! (\w^{(k)}\! -\!\gamma_1 (\nabla \TE(\w^{(k)})\!  +\! \v_1^{(k)} \!+\! \v_2^{(k)}))\!$}
   \STATE{$\mathbf{v_1}^{(k)} \leftarrow \mathbf{v_1}^{(k)} + \gamma_2 (2\w^{(k+1)} - \w^{(k)})$}
   \STATE{$\mathbf{v_2}^{(k)} \leftarrow \mathbf{v_2}^{(k)} + \gamma_2 (2\w^{(k+1)} - \w^{(k)})$}
   \STATE{$\v_1^{(k+1)} \leftarrow \v_1^{(k)} - \gamma_2 P_{S_{l,u}}(\frac{1}{\gamma_2} \v_1^{(k)})$}
   \STATE{$\v_2^{(k+1)} \leftarrow \v_2^{(k)} - \gamma_2 P_{S_{1}}(\frac{1}{\gamma_2} \v_2^{(k)})$}
   \STATE{$k \leftarrow k+1$}
  \ENDWHILE
  \RETURN $\w^{(k)}$ 
  \end{algorithmic} 
  \end{algorithm}

We use the primal-dual splitting method (PDS) to solve (\ref{eq:ours}).
We can use indicator functions $\iota_{S_s}$, $\iota_{S_{l,u}}$ and $\iota_{S_1}$, where 
\begin{align}
S_1 := \{\w \in \R^N \mid \mathbf1^\top \w = 1\},
\end{align}
to reformulate (\ref{eq:ours}) as
\begin{align}
  \label{eq:ours_pds}
  \begin{gathered}
   \min_{\w} \TE{(\w)} + \iota_{S_s}(\w) + \iota_{S_{l,u}}(\w) + \iota_{S_1}(\w).
  \end{gathered}
\end{align}

By defining $\v := [\v_1^\top\: \v_2^\top]^\top$ ($\v_1, \v_2 \in \R^{N}$), and $f_1, f_2, f_3, \A$ as
\begin{eqnarray} 
  \begin{gathered}
  \begin{split}
  \label{eq:pds2}
  f_1(\w) &:= \TE{(\w)},\\
  f_2(\w) &:= \iota_{S_s}(\w),\\
  f_3(\v) &:= \iota_{S_{l,u}}(\v_1) + \iota_{S_1}(\v_2),\\
  \A & := \begin{bmatrix}
          \I& \I
          \end{bmatrix}^\top,
  \end{split}
  \end{gathered}
\end{eqnarray}
where $\I \in \R^{N\times N}$ is an identity matrix, the problem in (\ref{eq:ours_pds}) is reduced to (\ref{eq:pds}).
Note that all of the tracking error measures introduced are $\beta$-Lipschitz continuous, differentiable convex functions, therefore satisfy the conditions on $f_1$ mentioned in Section \ref{sec:preliminary}-\ref{sec:tools}.
Although the PDS algorithm is guaranteed to converge when (\ref{eq:ours_pds}) is convex, this is not the case for our formulation because of the $\ell_0$-norm constraint. Although empirically the PDS algorithm converges most of the time, we gradually restrict the stepsizes $\gamma_1$ and $\gamma_2$ to stabilize the algorithm for nonconvex optimization. This is supported by studies of ADMM  and PDS algorithms for nonconvex cases, where stepsizes are diminished every iteration \cite{7178689,Guoyin2015,L0} or lowered in case of nonconvergence \cite{condat2015cadzow}.

When $S_s = S_0$, the proximity operator of $f_2: \prox_{\gamma\iota_{S_0}}$ is a projection\footnote{\setstretch{0.75}Strictly speaking, since $S_0$ is a nonconvex set, the projection onto this set cannot be defined as a one-to-one mapping, but fortunately, one of the projected points can be computed analytically as in (\ref{eq:l0_prox}).} onto the set $S_0$, which is as follows:
\begin{eqnarray} 
  \label{eq:l0_prox}
  \begin{gathered}
  \begin{split}
  [\prox_{\gamma\iota_{S_0}}(\z)]_i &= [P_{S_0}(\z)]_i\\ &= 
  \begin{cases}
  z_i, \:\mathrm{if} \ i \in \{(1),...(K_1)\},\\
  0,\:\mathrm{if} \ i \in \{(K_1+1)...(N)\},
  \end{cases}
\end{split}
\end{gathered}
\end{eqnarray}
where we denote the elements of $\z$ sorted in descending order in terms of their absolute values by $z_{(1)},...,z_{(N)}$, i.e., $|z_{(1)}|\geq |z_{(2)}|\geq...\geq|z_{(N)}|$.
In short, a projection onto the set $S_0$ can be calculated by leaving the $K_1$ largest absolute values and projecting the remaining elements of the vector to $0$.

When $S_s = S_{\w_0}$, the proximity operator (Algorithm \ref{al:ours1} line 2) is given by 
\begin{align}
  \begin{gathered}
\prox_{\gamma\iota_{S_{\w_0}}}(\z) = P_{S_{\w_0}}(\z) = \w_0 + P_{S_0}(\z-\w_0).
\end{gathered}
\end{align}

For $f_3$, the proximity operator of $\iota_{S_{l,u}}(\v_1)$ (line 5) is given by
\begin{align}
  \begin{gathered}
\prox_{\gamma\iota_{S_{l,u}}}(\z) = P_{S_{l,u}}(\z) = [\max\{l,\min\{z_i,u\}\}]_{1\leq i\leq N},
\end{gathered}
\end{align}
and the proximity operator of $\iota_{S_1}(\v_2)$ (line 6) is given by 
\begin{align}
  \begin{gathered}
\prox_{\gamma\iota_{S_{1}}}(\z) = P_{S_{1}}(\z) =  \z + \frac{1 - \mathbf1^\top\z}{\|\mathbf1\|_2^2}\mathbf1.
\end{gathered}
\end{align}

\subsection{Simultaneous Asset Selection and Capital Allocation}
When ETE is chosen as the tracking error measure and $S_s = S_0$, our method essentially solves the same problem as NNOMP-PGD (given that $u$ is large enough), and the difference is reduced to the optimization algorithm. 
We argue that the inherent ability of our algorithm to perform asset selection and capital allocation simultaneously endows it with superior index tracking performance—a claim substantiated in the experimental section (Section \ref{sec:experiments}-\ref{sec:track}).
NNOMP-PGD, in contrast, separates capital allocation into two steps: asset selection and capital allocation. While the impact of such a process on tracking accuracy is unclear, we suspect that even if this two-step procedure could, in theory, generate an optimal portfolio, it might not apply to NNOMP-PGD.

The NNOMP phase (responsible for asset selection) of the algorithm does not impose the sum-to-one constraint. Therefore, when a nonzero value is assigned to one of the elements of the portfolio, the value of the element is not restricted in any way. The assigned value directly affects the asset selection in the following iteration, since the updated portfolio is incorporated into the residual update of the current iteration. The updated residual is used in the next iteration to select the asset. 

Hence, asset selection may proceed under less-than-ideal conditions, lacking necessary constraints on values assigned in each iteration.
We hypothesize that this could impair the tracking performance and that simultaneous asset selection and capital allocation might prove beneficial.

\vspace{-3mm}
\subsection{Computational Complexity}
In this section, we discuss the computational complexity of the proposed algorithm.
The operations in our PDS-based algorithm (Algorithm \ref{al:ours1}) that potentially requires complex computations are the three proximity operations (projections), namely, $P_{S_s}$, $P_{S_{l,u}}$ and $P_{S_{1}}$, and $\nabla \TE$ involved in the line 2 of the algorithm. The projection $P_{S_s}$ is composed of two steps. The sorting process can be computed in the order of $O(N\log N)$ (we use the sort() function implemented in MATLAB, which uses quicksort), and the projection process in the order of $O(N)$. For the remaining $P_{S_{l,u}}$ and $P_{S_{1}}$, both can be computed in the order of $O(N)$. Gradient $\nabla \TE$ can be computed in the order of $O(NT)$.
Therefore, the overall computational complexity of the proposed algorithm per iteration can be simplified to $O(NT)$, since $T$ is generally large enough such that other operations can be disregarded.

\vspace{-2mm}
\section{Experiments}
\label{sec:experiments}
\subsection{Dataset and Settings}

\setlength{\aboverulesep}{0pt}
\setlength{\belowrulesep}{0pt}

\begin{table*}[t]
    
  \caption{The tracking performance measured in $\MDTE [\mathrm{bps}]$ and normalized accumulated returns (Ret.). Dataset: S\&P500, initial capital: $\$10000$. Note that LAIT is not included in Tables \ref{tb:sp} and \ref{tb:nas} due to the difficulties encountered in adjusting the sparsity to an exact value using LAIT.}
  \label{tb:sp}
  \begin{adjustbox}{center}
  \scalebox{0.9}{
  \begin{tabular}{cccccccccccccc}
    \toprule
    \multirow{3}{*}{Method} & \multirow{3}{*}{$K_2$} & \multicolumn{6}{c}{2012 - 2017} & \multicolumn{6}{c}{2017 - 2022} \\ 
    \cmidrule(lr){3-8} \cmidrule(lr){9-14}
                            & & \multicolumn{2}{c}{$K_1 = 40$} & \multicolumn{2}{c}{$K_1 = 60$} & \multicolumn{2}{c}{$K_1 = 80$} & \multicolumn{2}{c}{$K_1 = 40$} & \multicolumn{2}{c}{$K_1 = 60$} & \multicolumn{2}{c}{$K_1 = 80$} \\ 
    \cmidrule(lr){3-4} \cmidrule(lr){5-6} \cmidrule(lr){7-8} \cmidrule(lr){9-10} \cmidrule(lr){11-12} \cmidrule(lr){13-14}
                            & & MDTE & Ret. & MDTE & Ret. & MDTE & Ret. & MDTE & Ret. & MDTE & Ret. & MDTE & Ret. \\
    \midrule
    $\ell_0$-ADMM \cite{9843945}      & $-$    & $1.73 $ & $1.68 $ & $1.77 $ & $1.59 $ & $1.82 $ & $1.53 $ & $3.72 $ & $1.28 $ & $3.76 $ & $1.27 $ & $3.74 $ & $1.22 $\\
    NNOMP-PGD \cite{9775642}          & $-$    & $0.85 $ & $1.71 $ & $0.79 $ & $1.64 $ & $0.75 $ & $1.60 $ & $1.43 $ & $1.38 $ & $1.21 $ & $1.49 $ & $1.06 $ & $1.50 $\\\cmidrule(lr){1-2}
    Proposed[P, ETE]                  & $-$    & $0.74 $ & $1.75 $ & $0.64 $ & $1.71 $ & $0.48 $ & $1.73 $ & $1.13 $ & $1.43 $ & $0.79 $ & $1.46 $ & $0.68 $ & $1.47 $\\\cmidrule(lr){1-2}
    \multirow{3}{*}{Proposed[T, ETE]} & $K_1$  & $\mathbf{0.39} $ & $1.87 $ & $\mathbf{0.33} $ & $1.77 $ & $\mathbf{0.24} $ & $1.72 $ & $\mathbf{0.52} $ & $1.49 $ & $\mathbf{0.42} $ & $1.39 $ & $\mathbf{0.35} $ & $1.34 $\\
                                      & $K_1/2$& $0.47 $ & $1.93 $ & $0.37 $ & $1.81 $ & $0.31 $ & $1.81 $ & $0.61 $ & $1.62 $ & $0.55 $ & $1.48 $ & $0.41 $ & $1.41 $\\
                                      & $K_1/3$& $0.51 $ & $1.89 $ & $0.41 $ & $1.81 $ & $0.32 $ & $1.80 $ & $0.66 $ & $\mathbf{1.69} $ & $0.59 $ & $\mathbf{1.61} $ & $0.46 $ & $1.47 $\\\cmidrule(lr){1-2}
    Proposed[P, DR]                   & $-$    & $0.81 $ & $1.83 $ & $0.67 $ & $\mathbf{1.91} $ & $0.59 $ & $\mathbf{1.95} $ & $1.19 $ & $1.58 $ & $0.95 $ & $1.58 $ & $0.83 $ & $\mathbf{1.55} $\\\cmidrule(lr){1-2}
    \multirow{3}{*}{Proposed[T, DR]}  & $K_1$  & $0.42 $ & $1.92 $ & $0.35 $ & $1.76 $ & $0.31 $ & $1.76 $ & $0.64 $ & $1.49 $ & $0.46 $ & $1.36 $ & $0.40 $ & $1.36 $\\
                                      & $K_1/2$& $0.50 $ & $1.94 $ & $0.41 $ & $1.86 $ & $0.37 $ & $1.77 $ & $0.72 $ & $1.57 $ & $0.57 $ & $1.59 $ & $0.48 $ & $1.48 $\\
                                      & $K_1/3$& $0.52 $ & $\mathbf{1.99} $ & $0.44 $ & $1.89 $ & $0.39 $ & $1.86 $ & $0.84 $ & $1.67 $ & $0.59 $ & $1.50 $ & $0.56 $ & $1.57 $\\\midrule

    Benchmark                         & $-$    & $-$ & $1.38 $ & $-$ & $1.38 $ & $-$ & $1.38 $ & $-$ & $1.02 $ & $-$ & $1.02 $ & $-$ & $1.02 $ \\
    \bottomrule
  \end{tabular}
  }
\end{adjustbox}

\end{table*}

  \begin{table*}[t]
    \caption{The tracking performance measured in $\MDTE [\mathrm{bps}]$ and normalized accumulated returns (Ret.). Dataset: Russell3000, initial capital: $\$40000$. Note that LAIT is not included in Tables \ref{tb:sp} and \ref{tb:nas} due to the difficulties encountered in adjusting the sparsity to an exact value using LAIT.}
    \label{tb:nas}
    \begin{adjustbox}{center}
    \scalebox{0.9}{
      \begin{tabular}{cccccccccccccc}
      \toprule
      \multirow{3}{*}{Method} &\multirow{3}{*}{$K_2$}& \multicolumn{6}{c}{2010 - 2014}& \multicolumn{6}{c}{2015 - 2019} \\ 
      \cmidrule(lr){3-8} \cmidrule(lr){9-14}
      & &\multicolumn{2}{c}{$K_1 = 100$} &\multicolumn{2}{c}{$K_1 = 150$} &\multicolumn{2}{c}{$K_1 = 200$} &\multicolumn{2}{c}{$K_1 = 100$} &\multicolumn{2}{c}{$K_1 = 150$} &\multicolumn{2}{c}{$K_1 = 200$}\\ 
      \cmidrule(lr){3-4} \cmidrule(lr){5-6} \cmidrule(lr){7-8} \cmidrule(lr){9-10} \cmidrule(lr){11-12} \cmidrule(lr){13-14}
      & &MDTE&Ret.&MDTE&Ret.&MDTE&Ret.&MDTE&Ret.&MDTE&Ret.&MDTE&Ret.\\
      \midrule
      $\ell_0$-ADMM \cite{9843945}      & $-$    & $3.21 $ & $2.11 $ & $3.21 $ & $2.01 $ & $3.20 $ & $1.97 $ & $2.44 $ & $1.64 $ & $2.48 $ & $\mathbf{1.70} $ & $2.49 $ & $1.60 $\\
      NNOMP-PGD \cite{9775642}          & $-$    & $0.78 $ & $1.78 $ & $0.75 $ & $1.65 $ & $1.89 $ & $1.69 $ & $1.07 $ & $1.59 $ & $1.03 $ & $1.52 $ & $1.62 $ & $1.48 $\\\cmidrule(lr){1-2}
      Proposed[P, ETE]                  & $-$    & $0.83 $ & $2.28 $ & $0.65 $ & $\mathbf{2.20} $ & $0.54 $ & $1.92 $ & $0.88 $ & $1.53 $ & $0.68 $ & $1.47 $ & $0.58 $ & $1.57 $\\\cmidrule(lr){1-2}
      \multirow{3}{*}{Proposed[T, ETE]} & $K_1$  & $\mathbf{0.51} $ & $1.97 $ & $\mathbf{0.42} $ & $1.85 $ & $\mathbf{0.33} $ & $1.81 $ & $\mathbf{0.55} $ & $\mathbf{1.70} $ & $\mathbf{0.40} $ & $1.56 $ & $\mathbf{0.35} $ & $1.60 $\\
                                        & $K_1/2$& $0.56 $ & $1.92 $ & $0.47 $ & $1.80 $ & $0.37 $ & $1.79 $ & $0.61 $ & $1.68 $ & $0.46 $ & $1.64 $ & $0.40 $ & $1.60 $\\
                                        & $K_1/3$& $0.60 $ & $1.96 $ & $0.51 $ & $1.78 $ & $0.40 $ & $1.76 $ & $0.63 $ & $1.68 $ & $0.48 $ & $1.63 $ & $0.43 $ & $\mathbf{1.67} $\\\cmidrule(lr){1-2}
      Proposed[P, DR]                   & $-$    & $1.20 $ & $\mathbf{2.25} $ & $0.99 $ & $2.07 $ & $0.86 $ & $1.96 $ & $1.11 $ & $1.64 $ & $0.84 $ & $1.57 $ & $0.71 $ & $1.43 $\\\cmidrule(lr){1-2}
      \multirow{3}{*}{Proposed[T, DR]}  & $K_1$  & $0.58 $ & $1.99 $ & $0.44 $ & $2.00 $ & $0.36 $ & $1.90 $ & $0.62 $ & $1.55 $ & $0.49 $ & $1.55 $ & $0.41 $ & $1.46 $\\
                                        & $K_1/2$& $0.64 $ & $2.04 $ & $0.50 $ & $2.07 $ & $0.41 $ & $1.94 $ & $0.64 $ & $\mathbf{1.70} $ & $0.55 $ & $1.63 $ & $0.42 $ & $1.48 $\\
                                        & $K_1/3$& $0.71 $ & $2.02 $ & $0.51 $ & $2.00 $ & $0.44 $ & $\mathbf{1.99} $ & $0.69 $ & $1.60 $ & $0.57 $ & $1.57 $ & $0.44 $ & $1.48 $\\\midrule

      Benchmark                         & $-$    & $-$ & $1.37 $ & $-$ & $1.37 $ & $-$ & $1.37 $ & $-$ & $1.18 $ & $-$ & $1.18 $ & $-$ & $1.18 $ \\
      \bottomrule
      \end{tabular}
    }

    \end{adjustbox}
    
\end{table*}

\begin{table*}[t]
  \caption{The comparison between different initialization methods, tracking performance measured in $\text{MDTE} [\mathrm{bps}]$ and normalized accumulated returns (Ret.). Dataset: S\&P500, initial capital: $\$10000$, $K_1 = 40$.}
  \label{tb:initsp}
  \begin{adjustbox}{center}
  \scalebox{0.9}{
    \begin{tabular}{cccccccccccccc}
    \toprule
    \multirow{3}{*}{Method} & \multirow{3}{*}{$K_2$} & \multicolumn{6}{c}{S\&P500 (2012 - 2017)} & \multicolumn{6}{c}{Russell3000 (2010 - 2014)} \\ 
    \cmidrule(lr){3-8} \cmidrule(lr){9-14}
    & & \multicolumn{2}{c}{Init. A} & \multicolumn{2}{c}{Init. B} & \multicolumn{2}{c}{Init. C} & \multicolumn{2}{c}{Init. A} & \multicolumn{2}{c}{Init. B} & \multicolumn{2}{c}{Init. C} \\ 
    \cmidrule(lr){3-4} \cmidrule(lr){5-6} \cmidrule(lr){7-8} \cmidrule(lr){9-10} \cmidrule(lr){11-12} \cmidrule(lr){13-14}
    & & MDTE & Ret. & MDTE & Ret. & MDTE & Ret. & MDTE & Ret. & MDTE & Ret. & MDTE & Ret. \\
    \midrule
    Proposed[P, ETE]                  & $-$    & $0.74 $ & $1.75 $ & $0.58 $ & $2.03 $  & $0.61 $ & $1.93 $ & $1.13 $ & $1.43 $ & $0.76 $ & $1.65 $ & $0.78 $ & $1.78 $\\\cmidrule(lr){1-2}
    \multirow{3}{*}{Proposed[T, ETE]} & $K_1$  & $0.39 $ & $1.87 $ & $0.37 $ & $1.85 $  & $0.42 $ & $1.83 $ & $0.52 $ & $1.49 $ & $0.46 $ & $1.40 $ & $0.51 $ & $1.63 $\\
                                      & $K_1/2$& $0.47 $ & $1.93 $ & $0.43 $ & $1.93 $  & $0.52 $ & $1.89 $ & $0.61 $ & $1.62 $ & $0.57 $ & $1.46 $ & $0.67 $ & $1.63 $\\
                                      & $K_1/3$& $0.51 $ & $1.89 $ & $0.46 $ & $1.96 $  & $0.58 $ & $1.77 $ & $0.66 $ & $1.69 $ & $0.61 $ & $1.53 $ & $0.73 $ & $1.71 $\\\cmidrule(lr){1-2}
    Proposed[P, DR]                   & $-$    & $0.81 $ & $1.83 $ & $0.63 $ & $2.08 $  & $0.64 $ & $2.03 $ & $1.19 $ & $1.58 $ & $0.90 $ & $1.68 $ & $0.86 $ & $1.55 $\\\cmidrule(lr){1-2}
    \multirow{3}{*}{Proposed[T, DR]}  & $K_1$  & $0.42 $ & $1.92 $ & $0.35 $ & $1.86 $  & $0.44 $ & $1.88 $ & $0.64 $ & $1.49 $ & $0.50 $ & $1.49 $ & $0.63 $ & $1.56 $\\
                                      & $K_1/2$& $0.50 $ & $1.94 $ & $0.41 $ & $1.98 $  & $0.50 $ & $1.90 $ & $0.72 $ & $1.57 $ & $0.61 $ & $1.60 $ & $0.78 $ & $1.67 $\\
                                      & $K_1/3$& $0.52 $ & $1.99 $ & $0.46 $ & $1.95 $  & $0.54 $ & $1.99 $ & $0.84 $ & $1.67 $ & $0.62 $ & $1.65 $ & $0.79 $ & $1.60 $\\
    \midrule
    Benchmark & $-$ & $- $ & $1.38 $ & $- $ & $1.38 $ & $- $ & $1.38 $ & $- $ & $1.02 $ & $- $ & $1.02 $ & $- $ & $1.02 $ \\
    \bottomrule
    \end{tabular}
  }
\end{adjustbox}
\end{table*}

To test the performance of the index tracking methods, we adopt the rolling window scheme \cite{8064704,9775642}.
The first $T_{\mathrm{train}}$ time-frames are used to design the first portfolio, which will be used for out-of-sample testing in the next $T_{\mathrm{test}}$ time-frames. At the end of the testing period, we use the last $T_{\mathrm{train}}$ time-frames to design the next portfolio. We continue the training and testing cycle $n$ times. Therefore, a total of $T_{\mathrm{train}} + nT_{\mathrm{test}}$ time-frames are used for each experiment. The parameters are set to $n = 10$, $T_{\mathrm{train}} = 200$, and $T_{\mathrm{test}} = 100$.

We use the S\&P500 (September 2012 - August 2022) and Russell3000 (January 2010 - December 2019) index datasets, commonly used datasets for index tracking \cite{8064704,9775642} for the experiments.
Assets not covering the entire period are excluded, resulting in final datasets comprising 463 and 1624 assets for the S\&P500 and Russell3000, respectively.
The adjusted closing prices of the assets are used, and the return of an asset $j$ at time-frame $t$ is given by
\begin{align}
\X_{t,j}  = \frac{\mathrm{price}_{t,j} - \mathrm{price}_{t-1,j}}{\mathrm{price}_{t-1,j}}. 
\end{align}
We split both S\&P500 and Russell3000 index datasets into two, September 2012 - October 2017 and November 2017 - August 2022 (S\&P500), and January 2010 - October 2014 and March 2015 - December 2019 (Russell3000), respectively. Note that each time-frame represents a trading day, and because the stock exchange is closed on weekends, the dataset is not sampled regularly in the time direction. 

We measure how well the portfolio replicates the benchmark index by computing the magnitude of the daily tracking error (MDTE) defined as 
\begin{align}
  \begin{gathered}
    \MDTE = \frac{1}{nT_{\mathrm{test}}} \|\diag(\X\W)-\r^\b\|_2,
\end{gathered}
\end{align}
where $\diag(\cdot)$ indicates a vector consisting of the diagonal elements of a given matrix. $\W \in \R^{N\times nT_{\mathrm{test}}}$ is a matrix where the portfolio designed at the end of a training period is stacked for the following test period, so that column $t$ contains the portfolio used in time-frame $t$.
Note that the MDTE value is presented in basis points ($1 \mathrm{bps} = 10^{-4}$).
Since we could not obtain the asset-wise weights from the index sponsors, we use a uniform portfolio $\b$ as a benchmark index, where all of the capital is allocated evenly across all $N$ assets. The benchmark index return is given by $\r^\b = \X\b$.

Furthermore, we conduct a simulation of actual investments made based on the computed portfolios. We simulate rebalancing based on a new portfolio at the start of every test period. The acquired returns are reinvested. We apply a transaction costs model common in the U.S. markets: the cost per transaction is $\$0.005\times v$ with a minimum cost of $\$1$. The Ret. columns of Tables \ref{tb:sp} and \ref{tb:nas} show the normalized accumulated return at the end of the entire testing period.

As for the comparison methods, in addition to the aforementioned NNOMP-PGD \cite{9775642} and LAIT \cite{8064704}, the state-of-the-art sparse index tracking methods, we also compare our method with $\ell_0$-ADMM (Alternating Direction Method of Multipliers) \cite{9843945}. 
Despite how $\ell_0$-ADMM focuses on optimizing between return and risk, making it distinct from typical index tracking approaches, its implementation of an $\ell_0$-norm constraint mirrors that of NNOMP-PGD and our proposed method, warranting its inclusion in our comparison.
Regarding our methods, portfolio-sparse methods (Proposed[P, TE]) allocate the capital from scratch every rebalancing, in other words, $S_s = S_0$. For turnover-sparse methods (Proposed[T, TE]), the turnovers are sparsified instead of the portfolio ($S_s = S_{\w_0}$) with $K_2$ set to various values. Note that turnover-sparse methods adopt $S_s = S_0$ for the first training period since there is no previous portfolio to refer to.
We also evaluate the performance of different tracking error measures (ETE and DR).

Regarding the other settings of the experiment, the stopping criterion is set to 
\begin{align}
\frac{\|\w^{(k)} - \w^{(k-1)}\|_2}{\|\w^{(k-1)}\|_2} \leq 1.0 \times 10^{-5}. 
\end{align}
For the box constraint in (\ref{eq:ours}), the lower and upper bounds are set to $l = 0, u = 4/K_1$. 
The stepsizes $\gamma_1$ and $\gamma_2$ of the proposed algorithm are first set to sufficiently meet the conditions mentioned in Section \ref{sec:preliminary}-\ref{sec:tools}, and then multiplied by $0.999$ every iteration.
\vspace{-3mm}

\begin{figure}
\begin{center}
    \begin{center}
    \includegraphics[clip, width=0.5\textwidth]{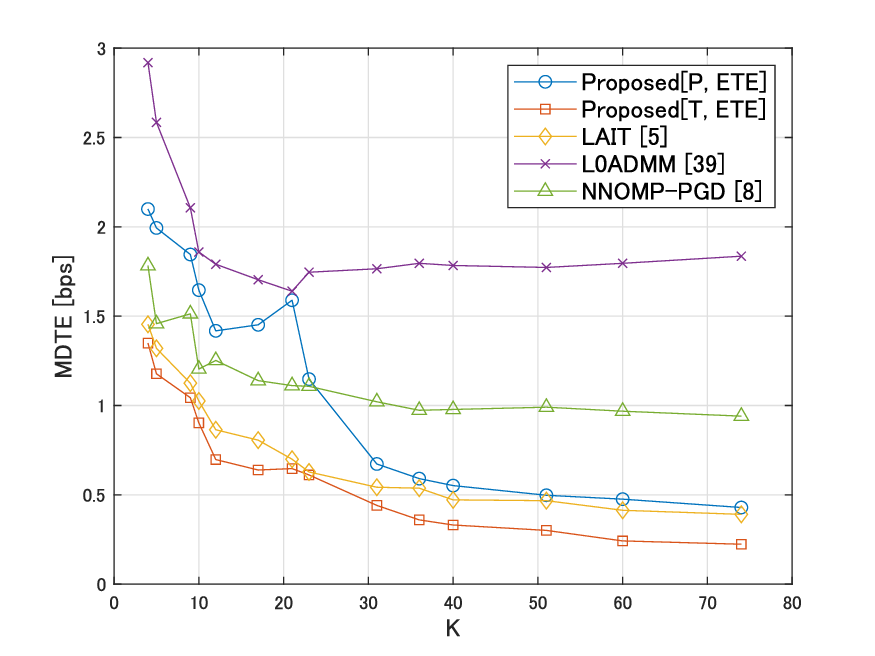}
    \end{center}
    \vspace{-2mm}
\caption{The graph of $\MDTE [\mathrm{bps}]$ across different sparsity on S\&P500, 2012 - 2017. The vertical axis indicates $\MDTE [\mathrm{bps}]$ and the horizontal axis indicates the sparsity. To avoid parameter tuning on $\lambda$ (which controls sparsity in LAIT), we fixed the value of $\lambda$ per data point. Therefore, the exact sparsity of per training period differs. The sparsity of Proposed[P, ETE] ($K_1$), NNOMP-PGD and $\ell_0$-ADMM is adjusted to be the same as that of LAIT. As for Proposed[T, ETE], $K_2 = K_1$. Parameter $K$ in the graph indicates the sparsity of the portfolio at $n = 10$.} 
\label{fig:MDTE}
\end{center}
\vspace{-4mm}
\end{figure}

\begin{figure}
  \begin{center}
  
      \begin{center}
      \includegraphics[clip, width=0.5\textwidth]{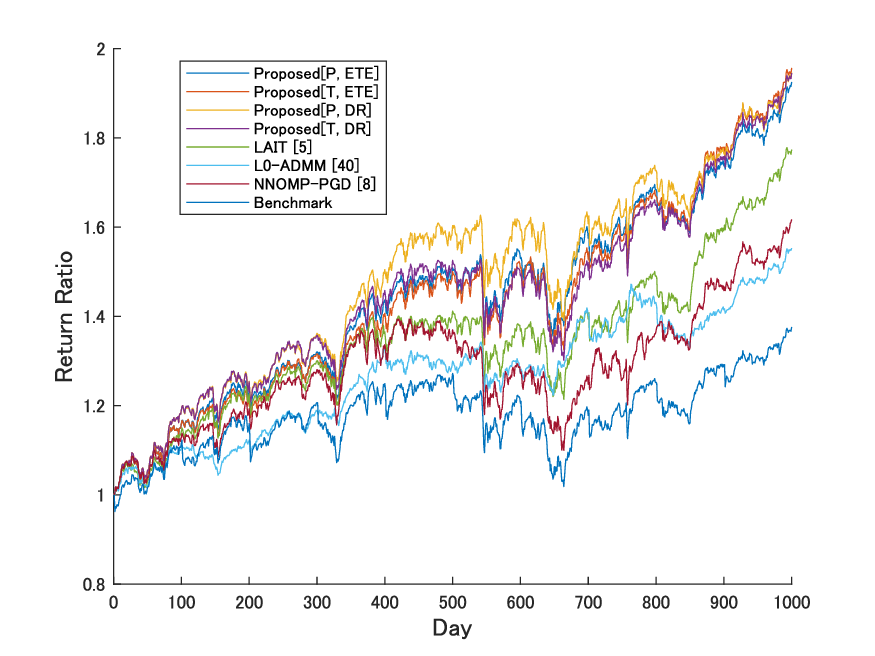}
      \end{center}

      \vspace{-2mm}
  \caption{The graph of the investment simulation on S\&P500, 2012 - 2017. The vertical axis indicates the normalized accumulated return and the horizontal axis indicates the sparsity of the portfolio ($K_1,K=40$ and $K_2 = K_1/3$). The graph indicates an investment simulation based on one of the data points of Fig. \ref{fig:MDTE}. The initial capital is $\$10000$.} 
  \label{fig:investment}
  \end{center}
  \vspace{-5mm}
  \end{figure}

\subsection{Results}

\subsubsection{Tracking Performance}
\label{sec:track}
Across all evaluated settings, our proposed method surpassed both $\ell_0$-ADMM and NNOMP-PGD in performance on the S\&P500 and Russell3000 datasets, as detailed in Tables \ref{tb:sp} and \ref{tb:nas} (MDTE).
This supports our hypothesis that concurrent asset selection and capital allocation can be advantageous. 
Both Proposed[P, ETE] and NNOMP-PGD strategies utilize similar $\ell_0$-norm constraints and tracking error measures. The primary difference between them lies in the algorithms used for solving these constraints.
Note that LAIT is not included in Tables \ref{tb:sp} and \ref{tb:nas} due to the difficulties encountered in adjusting the sparsity to an exact value using LAIT. 
Regarding LAIT (Fig. \ref{fig:MDTE}), while the turnover-sparse (Proposed[T, ETE]) method achieved better tracking performance, the portfolio-sparse (Proposed[P, ETE]) method slightly underperformed LAIT. The superior performance of LAIT likely stems from its guaranteed convergence to an optimal solution. The proposed portfolio-sparse method's slight underperformance should be considered as a trade-off between its ability to directly control the sparsity of the portfolio.

Furthermore, comparing the tracking performance of portfolio-sparse (Proposed[P, ETE]) methods and turnover-sparse (Proposed[T, ETE]) methods, we can see that turnover-sparse methods always performed better. We believe this is because portfolios constructed using turnover-sparse methods are composed of more nonzero weights compared to those of portfolio-sparse methods. Because turnover-sparse methods only enforce the sparseness of the turnover, the sparsity of the portfolio itself is not considered. Therefore, as the simulation progresses, the portfolio gradually becomes fuller. A fuller portfolio can replicate the target index with more nonzero weights, which should affect the tracking performance positively. Turnover-sparse methods with $K_2 = K_1$ tracked the index more effectively than other $K_2$ values did, for the same reason, because larger $K_2$ values densify the portfolio faster than smaller $K_2$ values. Similarly, a larger $K_1$ performed better for portfolio-sparse methods, since portfolios constructed from larger $K_1$ values have more nonzero weights.

\vspace{-6mm}
\subsubsection{Return Accumulation}

Our proposed methods outperformed the benchmark index and other comparative methods in terms of return accumulation across most settings, as detailed in Tables \ref{tb:sp} and \ref{tb:nas} (Ret.).
It is also clear that sparse portfolios are much more efficient compared to benchmark (full) portfolios (Fig. \ref{fig:investment}). 
However, regarding comparisons between portfolio-sparse and turnover-sparse methods and the various tracking measures, we acquired varying results depending on the dataset and period. 

First, when we applied the empirical tracking error (ETE) as the tracking measure, turnover-sparse methods (Proposed[T, ETE]) generally performed better than portfolio-sparse methods (Proposed[P, ETE]). This seems intuitive since turnover-sparse methods sparsify the turnover, which directly affects the transaction costs. In the worst-case rebalancing scenario, portfolios constructed by portfolio-sparse methods sell all assets in possesion and newly buy completely different assets. In this case, the transaction cost would be $\$2K_1$ (assuming the minimum cost is dominant). In comparison, the worst-case scenario for portfolios constructed by turnover-sparse methods is always $K_2$. 

However, results became more varied and inconsistent when applying the downside risk (DR) as the tracking measure.
At present, we lack a definitive explanation for these inconsistencies, though we speculate they may stem from the challenges associated with the nonconvex nature of the formulation, a topic we further discuss in Section \ref{sec:experiments}-\ref{sec:init}.
In addition to the above behavior, DR-applied methods performed moderately in comparison to ETE-applied methods. We expected DR to accumulate more returns than ETE, but there are some results where ETE outperformed the DR counterpart. 
While DR is designed to trade tracking accuracy for potentially higher returns, our results indicate that this trade-off does not consistently yield the expected benefits.

\vspace{-5mm}
\subsubsection{Initialization}
\label{sec:init}

Given the non-convex nature of our formulation, convergence cannot be guaranteed by our proposed algorithm. Therefore, the method of initializing the parameter $\w$ may potentially impact the results.
To understand how initialization impacts our algorithm's performance, we evaluated three distinct initialization strategies: Init. A: $\w = \mathbf{0}$, the original initialization adopted in our algorithm (Algorithm \ref{al:ours1}). Init. B: initialize all elements of $\w$ to $1/N$. Init. C: initialize $\w$ as $\w = \w_0$ ($\w = \mathbf{0}$ for the first training period).

The varied outcomes, as detailed in Table \ref{tb:initsp}, underscore the effect of the chosen initialization method on the algorithm's performance.
Given the inconsistency of results across different datasets, identifying a universally superior initialization method proved challenging.
It is noteworthy, however, that our method consistently outperformed NNOMP-PGD across the majority of settings, irrespective of the initialization technique employed.

\begin{figure}
  \begin{center}
      \begin{center}
      \includegraphics[clip, width=0.5\textwidth]{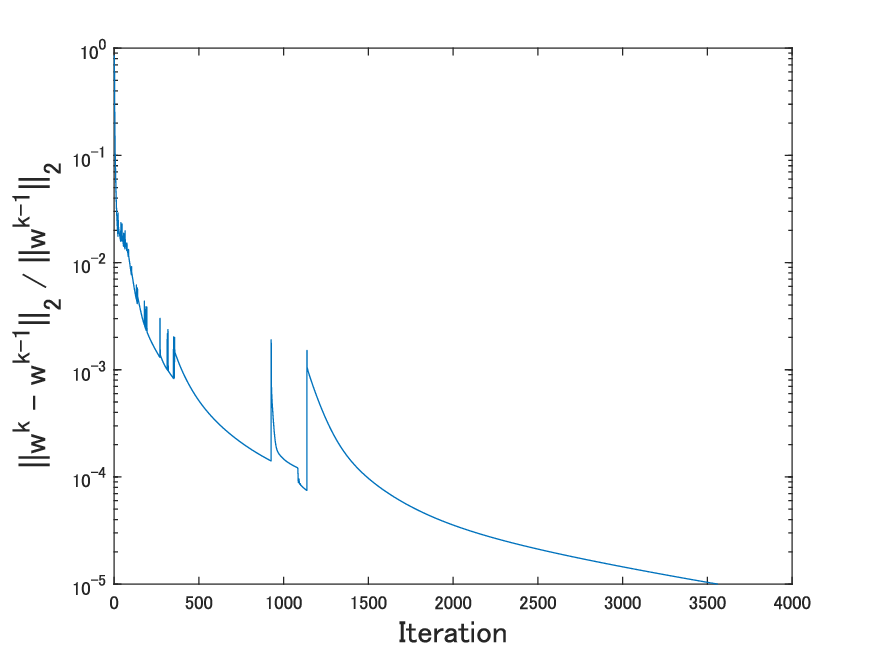}
      \end{center}
  
  \caption{The graph of the proposed algorithm's ([P,ETE]) convergence behavior on the S\&P500 (2012 - 2017) dataset ($K_1 = 40$, Init. A). The vertical axis indicates $\frac{\|\w^{(k)} - \w^{(k-1)}\|_2}{\|\w^{(k-1)}\|_2}$ and the horizontal axis indicates the number of iterations.} 
  \label{fig:conv}
  \end{center}
  \vspace{-6mm}
  \end{figure}

Furthermore, we present the convergence behavior of our algorithm in Fig. \ref{fig:conv}. 
Our algorithm converged in a similar manner across all experimental settings. 
Despite lacking a formal convergence guarantee, the observed convergence behavior and compelling tracking performance allow us to conclude that our method is capable of generating competitive portfolios, on par with methods like LAIT that do guarantee convergence.

\vspace{-5mm}
\subsubsection{Practical Application}
Based on the previous results and discussions, we would like to discuss the possibility of applying our methods to real-world investment situations.
First, the data and variables used in our methods to construct the portfolios are available.
Therefore, the main concern would be the choice of portfolio or turnover sparsity and tracking error measures. 
Despite the absence of a universally superior method across datasets, a feasible strategy for choosing a method in practice involves selecting the one that yielded the highest returns in the most recent data analysis.
In the same manner, parameters $K_1$ and $K_2$ can also be decided.

One of the pros of a sparse portfolio is that it can avoid illiquid assets.
However, extensive application of turnover-sparse methods may result in dense or full portfolios, as discussed in Section \ref{sec:experiments}-\ref{sec:track}. In practical scenarios, maintaining an optimally sparse portfolio can involve periodically resetting the portfolio using the portfolio-sparse method, particularly to exclude increasingly illiquid assets.

Although we did not explicitly discuss this in our paper, the length of $T_{\mathrm{train}}$ and $T_{\mathrm{test}}$ are also parameters. The two should be decided by simulating past data, similar to the other parameters.

\vspace{-6mm}
\subsubsection{Summary of the Experimental Results}
 Through extensive numerical experiments on S\&P500 and Russell3000 datasets, we:

 \begin{itemize}
    \item Confirmed our hypothesis that simultaneous asset selection and capital allocation can be beneficial in terms of tracking accuracy.
    \item Presented the merits of turnover sparsity in both index tracking and wealth accumulation.
    \item Investigated the impact of different tracking error measures and associated parameters on portfolio performance.
    \item Analyzed the convergence behavior and how various initialization methods influence the effectiveness of the proposed algorithm.
    \item Explored the practical applicability of our method in real-world investment scenarios.
 \end{itemize}

\vspace{-5mm}
\section{Concluding Remarks}

In this paper, we proposed a sparse index tracking method that addressed both asset selection and capital allocation simultaneously, enhancing tracking performance compared to the conventional method that handled these two aspects separately. Furthermore, the proposed formulation was generalized to allow the choice between 1) portfolio sparsity and turnover sparsity constraints, both enforced by an $\ell_0$-norm constraint and 2) various tracking error measures aimed at enhancing return accumulation performance. Superior results were demonstrated through experiments on the S\&P500 and Russell3000 index datasets, where we achieved state-of-the-art performance compared to the conventional method that incorporated an $\ell_0$-norm constraint. We also examined and discussed the impacts of different tracking measures and initializations.



\bibliographystyle{_IEEEtrans}

\end{document}